# Piezoresistivity and Strain-induced Band Gap Tuning in Atomically Thin MoS$_2$

Sajedeh Manzeli, Adrien Allain, Amirhossein Ghadimi, Andras Kis[*]

*Electrical Engineering Institute, École Polytechnique Fédérale de Lausanne (EPFL), CH-1015 Lausanne, Switzerland*

*Correspondence should be addressed to: Andras Kis, andras.kis@epfl.ch

**ABSTRACT**

Continuous tuning of material properties is highly desirable for a wide range of applications, with strain engineering being an interesting way of achieving it. The tuning range is however limited in conventional bulk materials which can suffer from plasticity and low fracture limit due to the presence of defects and dislocations. Atomically thin membranes such as MoS$_2$ on the other hand exhibit high Young's modulus and fracture strength which makes them viable candidates for modifying their properties *via* strain. The bandgap of MoS$_2$ is highly strain-tunable which results in the modulation of its electrical conductivity and manifests itself as the piezoresistive effect while a piezoelectric effect was also observed in odd-layered MoS$_2$ with broken inversion symmetry. This coupling between electrical and mechanical properties makes MoS$_2$ a very promising material for nanoelectromechanical systems (NEMS). Here we incorporate monolayer, bilayer and trilayer MoS$_2$ in a nanoelectromechanical membrane configuration. We detect strain-induced band gap tuning *via* electrical conductivity measurements and demonstrate the emergence of the piezoresistive effect in MoS$_2$. Finite element method (FEM) simulations are used to quantify the band gap change and to obtain a comprehensive picture of the spatially varying bandgap profile on the membrane. The piezoresistive gauge factor is calculated to be −148 ± 19, −224 ± 19 and −43.5 ± 11 for monolayer, bilayer and trilayer MoS$_2$ respectively which is comparable to state-of-the-art silicon strain sensors and two orders of magnitude higher than in strain sensors based on suspended graphene. Controllable modulation of resistivity in 2D nanomaterials using strain-induced bandgap tuning offers a novel approach for implementing an important class of NEMS transducers, flexible and wearable electronics, tuneable photovoltaics and photodetection.

MoS$_2$ consists of vertically stacked, weakly interacting layers held together by van der Waals interaction and is a typical material from the transition metal dichalcogenide (TMDC) family[1]. While few-layer TMDCs are indirect bandgap semiconductors, they become direct gap semiconductors in their monolayer form.[2–6] The demonstration of the first monolayer MoS$_2$-based transistor[7] opened the way to fundamental studies and practical applications based on electrical transport in mesoscopic TMDC materials and enabled fabrication of high performance electronic and optoelectronic devices based on these materials[8]. From the mechanical point of view, MoS$_2$ benefits from its atomic scale thickness, ultralow weight and low intrinsic mechanical dissipation making it interesting for the realization of nanoresonators.[9] It has a high Young's modulus of ~270 GPa,[10] can sustain in-plane strain levels as high as 11%[10] which puts it in the category of ultrastrong materials. It can also avoid inelastic relaxation due to its high elastic strain limit[11]. All these features make $MoS_2$ and other TMDCs in general interesting for strain engineering and have motivated numerous theoretical studies[12,13,14,15,16,17], showing for example that under small compressive strains (< 2%) the bandgap is expected to increase[17]. Under tensile strain, the bandgap of monolayer and bilayer $MoS_2$ should be reduced and result in a semiconductor to metal transition for strain levels as high as 10% for monolayer and 6% for bilayer $MoS_2$.[16,17,18,19] Moreover, in the case of monolayer $MoS_2$, the indirect gap is only slightly higher in energy compared to the direct gap[17,20] and is more sensitive to strain, thus direct to indirect gap transition is expected under moderate strains.[13,17,18] The strain induced bandgap modulation gives rise to a piezoresistive effect, in which a change in resistivity of the material is observed during mechanical deformation, as was previously reported in the case of bilayer $MoS_2$ deposited on flexible substrates.[21] In addition to piezoresistivity, odd-layered ultrathin $MoS_2$ was also shown to exhibit the piezoelectric effect.[21,22]

While experimental reports on the strain-induced modification of the bandgap under tensile strain[12] and inhomogeneous local strain[23] have been published, they all relied on optical measurements. He et. al. reported an exciton redshift at a rate of ~70 meV/% strain for single-layer and at a larger rate for bilayer $MoS_2$[12]. Raman spectroscopy revealed the effect of strain on the vibrational modes and the strain-induced symmetry breaking[24]. Local strain engineering was also shown to result in the funnel effect[23]. The use of photoluminescence spectroscopy in most of these studies however restricts these studies to mono and bilayer $MoS_2$ and there is a lack of experimental information on the influence of strain-induced bandgap changes on electrical properties of $MoS_2$ which is needed to assess the potential of this class of materials as building blocks for NEMS devices such as self-sensing resonators or strain sensors[25].

Here, we investigate the effect of mechanical strain on the electrical conductivity of suspended $MoS_2$ membranes. We apply mechanical strain using an atomic force microscope tip while simultaneously carrying out electrical measurements, allowing us to detect the strain-induced bandgap modulation through its influence on the electrical conductivity. The device fabrication starts by mechanical exfoliation of mono- bi- and trilayer $MoS_2$ onto intrinsic, undoped Si substrates covered with a 270 nm thick $SiO_2$ layer. The use of intrinsic Si and the absence of gate electrodes minimizes the effect of capacitive coupling between the $MoS_2$ membrane and the substrate. Electrodes are fabricated using standard e-beam lithography and the 2D semiconductor is suspended by etching away a portion of the underlying $SiO_2$. Figure 1a shows a typical suspended $MoS_2$ device made from a single $MoS_2$ flake with electrical contacts clamping the suspended membrane.

In order to probe the electromechanical response of the suspended $MoS_2$ membrane, after the $MoS_2$ membrane is imaged and located using AFM in the AC mode, we position the AFM tip in the center of the membrane and deform it while applying a DC bias voltage ($V_{ds}$) between source and drain contacts. The drain current ($I_{ds}$) and membrane deformation $\delta_{mem}$ are recorded in-situ in a setup schematically depicted on Figure 1b. During a nanoindentation measurement, the piezoscanner displaces the AFM probe in the vertical direction with a controlled speed. Figure 2a shows the electrical response to the nanoindentation of device N°1 (monolayer, see Supplementary Section 3). The tip of the probe touches the membrane and starts to deform it, resulting in increasing deflection of the AFM probe $\delta_{probe}$. Once a predefined deflection is attained, the probe is retracted. At the same time, we record the drain current ($I_{ds}$) under a bias voltage ($V_{ds}$ = 200 mV), shown in the lower panel of Figure 2a. Both the electrical and mechanical response are reproducible over extension and retraction cycles of the AFM piezo scanner motion, indicating that the membrane is deformed in the elastic regime and that there is no slippage of the membrane under the metal contacts (Supplementary Figure S7). In addition, the reproducibility of current measurements shows that the contact interface between $MoS_2$ and the electrodes is not degraded during the nanoelectromechanical testing.

Concurrent measurements of the current and tip deflection allow us to observe the effect of deformation and confirm the mechanical origin of the current modulation. In the case of measurements on the monolayer $MoS_2$ device presented on Figure 2a, while the membrane is in the relaxed state, the current remains at a constant value of 470 pA. It starts to increase as soon as the membrane begins to deform, reaching a value of ~800 pA at maximum deformation. During the probe retraction cycle, current follows the opposite trend and returns to its pre-deformation value as the tip is fully retracted. The deflection of the membrane $\delta_{mem}$ at the center (right under the AFM tip) is related to the probe deflection $\delta_{probe}$ and vertical position of the piezo-scanner $z_{piezo}$ by $z_{piezo} = \delta_{mem} + \delta_{probe}$[26] (Figure 1b). This allows us to plot the current $I_{ds}$ as a function of $\delta_{mem}$ in Figure 2b. As shown in Figure 2b, the current $I_{ds}$ increases with the increased deformation of the membrane indicating the modulation of resistance due to the applied deformation.

In a different set of measurements, the output characteristics of the device is compared in the relaxed state and under constant deformation (Figure 2c). The current is systematically higher under a deformation of $\delta_{mem}$ = 33 nm than in the absence of deformation (black curve). Both curves are linear and symmetric, indicating that the piezotronic effect[21] due to a change in Schottky barrier height by piezoelectric polarization charges is negligible in this case and that the device response is dominated by the piezoresistive effect.

We have performed the same set of electromechanical measurements on 6 monolayers, 3 bilayer and 3 trilayer $MoS_2$ devices of various widths (85 nm − 6 µm), lengths (570 nm − 1.4 µm) and aspect ratios (length/width = 0.17 – 13), presented in Supplementary Section 3. In all cases, we eventually deformed the $MoS_2$ membrane up to mechanical failure (Supplementary Figure S4) with the current increasing with increasing deformation in all the cases.

This observed piezoresistive behavior can be understood in terms of band gap reduction under tensile strain[12,13,23]. In the sub-threshold regime and at room temperature, thermally activated transport dominates and the electrical current is carried by electrons thermally excited into the conduction band. The conductivity is then expressed as[27]:

$$\sigma = \sigma_0 \exp\left[-\frac{E_C - E_F}{k_B T}\right] \quad (1)$$

where $\sigma_0$ is the minimum conductivity defined by the hopping distance, $E_C$ is the conduction band edge, $E_F$ is the Fermi energy, $k_B$ is the Boltzmann constant and $T$ the temperature. Assuming a symmetric reduction of bandgap under strain[17,23,28,29,30,31,32], the conduction band would be shifted to lower energies while the valence band edge would be shifted to higher energies by the same amount. For small strains (up to 7% in our case), the bandgap is expected to change linearly with strain[12,17,18,30,33] and the conductivity can be written as (see Supplementary Section 5 for derivation):

$$\sigma_{def} = \sigma_{rel} \exp\left[-\frac{\varepsilon}{2k_B T}\frac{\partial E_g}{\partial \varepsilon}\right] \qquad (2)$$

$\sigma_{def}$ and $\sigma_{rel}$ are respectively the conductance of the membrane in the deformed and relaxed state, $\varepsilon$ is strain and $\partial E_g/\partial \varepsilon$ is the rate of bandgap change with strain. For negative values of $\partial E_g/\partial \varepsilon$, Eq. 2 predicts increasing conductivity.

The strain distribution induced in our MoS$_2$ membranes during the nanoindentation experiment is however not uniform, motivating the use of finite element modelling (FEM) for extracting the rate of band gap change $\partial E_g/\partial \varepsilon$ from our measurements. Figure 3a shows an AFM image of an electromechanical device based on a suspended trilayer MoS$_2$ membrane (device N°11, see Supplementary Section 3). Using FEM, we calculate the total conductance of the membrane as a function of $\delta_{mem}$, for a range of $\partial E_g/\partial \varepsilon$ and different values of contact resistance $R_C$. Figure 3b shows simulated resistance as function of membrane deflection for $\partial E_g/\partial \varepsilon$ in the range between 0 and −100 meV/% strain and $R_C$ = 2 MΩ. Comparing simulations and measurement results, it is possible to select among the simulated values of $\partial E_g/\partial \varepsilon$ and $R_C$ a pair of parameters that gives the smallest sum of the squared difference between the observed and simulated values, resulting in this particular case in $\partial E_g/\partial \varepsilon$ = − 21 meV/% strain and $R_C$ = 2 MΩ. The extracted value of contact resistance is in line with previously reported values[34] and corresponds to ~8% of the total device resistance in its relaxed state. Since our devices are in the sub-threshold regime due to the absence of gating, the resistance of the semiconducting channel is dominant and thus the piezoresistive behaviour of the channel is not masked by the effect of the in-series contact resistance.

In Figure 3c we show FEM simulation of the spatial distribution of bandgap change $\Delta E_g$ under an inhomogeneous strain resulting from a membrane deformation at midpoint equal to $\delta_{mem}$ = 75 nm. A spatially inhomogeneous strain field generates a spatially varying bandgap in an initially homogenous atomically thin membrane. The profile of the bandgap change $\Delta E_g$ along the dashed line in Figure 3c is shown in Figure 3d indicating that areas near the tip are experiencing more deflection and thus more strain and the largest change of band gap.

Measured and simulated curves for representative mono-, bi- and trilayer devices (samples N°1, N°7 and N°11, see Supplementary Section 3) are shown on Figure 4a-c.

Figure 4d depicts the calculated |$\partial E_g/\partial \varepsilon$| for all devices (for the geometry of each device see Supplementary Section 3). The error bars are calculated considering the uncertainty on the input parameters. We find that the bandgap is being tuned at rates of −77.3 ± 10 meV/% strain, −116.7 ± 10 meV/% strain and −22.7 ± 6 meV/% strain for monolayer, bilayer and trilayer MoS$_2$ respectively, in excellent agreement with theoretical predictions and optical measurements.[12,18,23,35]

The bandgap tuning rate is higher in bilayer than in monolayer devices. This result is consistent with previous theoretical and experimental reports[12,18,36]. The orbital contributions of the band-edge states and their hybridizations are different between the monolayer and bilayer MoS$_2$ and are differently affected by strain, thus leading to different rates of band gap change. Under tensile strain in the planar direction, the in-plane orbital hybridizations are modified. Due to the Poisson effect, the distance between atomic layers is reduced which will influence the out-of-plane orbital hybridizations as well. Monolayer $MoS_2$ consists of only one Mo plane, therefore the Mo $d_z^2$ orbitals, which are along the out-of-plane direction, are not affected by strain. On the other hand, in bilayer $MoS_2$ with two Mo planes, the Poisson contraction leads to a stronger interaction between Mo $d_z^2$ orbitals of the two Mo planes. Because of the higher sensitivity of the out of plane orbitals to the strain, the indirect band gap of bilayer $MoS_2$ shows a higher $|\partial E_g/\partial \varepsilon|$.[19] More in-depth theoretical investigations are required to explain the effect of strain on orbital interactions which change the bandgap of trilyer $MoS_2$.

Using finite element modelling, we can also extract the piezoresistive gauge factor (GF), defined as $GF = (\Delta R/R_0)/\varepsilon$ where $R_0$ is the total resistance of the unstrained $MoS_2$ channel and $\Delta R$ the resistance change under strain $\varepsilon$. Although the strain distribution in our membranes is not uniform, it varies continuously and smoothly and the strain experienced by an infinitesimally small element in the model is uniform. According to Eq. (2), the resistance $r$ of a finite element under strain $\varepsilon$ can be written as $r = r_0 \exp(\alpha\varepsilon)$ where $r_0$ is the resistance in the absence of strain and $\alpha = [1/(2k_BT)]\times[\partial E_g/\partial \varepsilon]$. For small strains we have that the gauge factor $GF \approx \alpha$ (Supplementary Section 8). Using the values of $\partial E_g/\partial \varepsilon$ found for $MoS_2$, we find that the piezoresistance gauge factor for monolayer, bilayer and trilayer $MoS_2$ is $-148 \pm 19$, $-224 \pm 19$ and $-43.5 \pm 11$ respectively, with the negative sign indicating decreasing resistivity with increasing strain. This is in contrast to graphene, were the application of strain results in decreasing Fermi velocity and reduced mobility, resulting in increasing resistivity[37]. Moreover, it is interesting to note that the piezoresistive gauge factor is highest in bilayer $MoS_2$ which is due to the higher sensitivity of the bandgap to strain. The gauge factors of monolayer and bilayer $MoS_2$ measured in our experiment are two orders of magnitude higher than in graphene strain sensors (~2)[37,38] and comparable to state-of-the-art silicon strain sensors (~200)[39]. Silicon however has a much lower fracture strain (0.7%)[40] than $MoS_2$ (as high as 11%)[10], implying that the latter would be more suitable for strain measurements on curved surfaces and highly deformable objects such as biological tissue. The large piezoresistive coefficient together with the atomic scale thickness also makes $MoS_2$ suitable for fabrication of self-sensing nanoelectromechanical systems and transparent strain gauges. Chemical doping[41] could be used in future to reduce the power dissipation in practical devices due to the relatively high device resistance, currently in the M$\Omega$ range because of the absence of gating.

In conclusion, we have demonstrated strain-induced tuning of the bandgap and electrical resistance of atomically thin layers of $MoS_2$. A finite element method analysis was developed to model the experimental observations. We show that the bandgap of $MoS_2$ decreases under mechanical strain and that $MoS_2$ has a piezoresistive gauge factor comparable to state-of-the-art silicon strain sensors. The developed methodology is generally applicable to other transition metal dichalcogenide semiconductors. Our study reveals that similarly to CMOS devices[42], strain, which can be easily controlled through the device fabrication process, is an effective agent to alter electronic transport properties in $MoS_2$, enabling its efficient implementation as piezoresistive transducer elements for emerging NEMS sensors.

## METHODS

**Fabrication of suspended MoS₂ devices clamped at both ends.** MoS$_2$ flakes were mechanically exfoliated onto an intrinsic Si substrate with 270 nm of SiO$_2$. The substrate is imaged using an optical microscope (Olympus BX51M) equipped with a color camera. We have previously established the correlation between the optical contrast and thickness as measured by AFM for a number of dichalcogenide materials, including MoS$_2$[43]. Mono-, bi- and trilayer MoS$_2$ flakes were optically detected with thickness confirmed using AFM topography imaging. Contacts are prepared using standard e-beam lithography and e-beam evaporation of Cr/Au (2nm/60nm) and lift-off in acetone. The devices were annealed at 200 °C in order to remove resist residue and decrease the contact resistance The suspension of the channel is achieved by etching away the underlying SiO$_2$ using buffered hydrofluoric acid (BOE 7:1). In order to prevent the MoS$_2$ membranes from collapsing due to the surface tension during the drying process, the suspended MoS$_2$ was released in a critical point drier (CPD). Prior to measurements, the suspended devices were annealed in vacuum (6.7 × 10$^{-6}$ mbar) at 150$^0$C for 20 hours in order to remove residues and adsorbates from both surfaces of the membrane.

**Electromechanical measurements.** Nanomechanical testing is performed using a commercially available AFM (Asylum Research Cypher). We use Mikromasch HQ probes (Model NSC35/AL BS). The photodetector is calibrated by performing nanoindentation on the SiO$_2$ substrate. The calibration curve is presented in Supplementary Figure S2. In addition to the calibration the spring constant of each cantilever was extracted prior to electromechanical measurements using the thermal noise method[44]. Current measurements are carried out using a Stanford Research System SR570 current amplifier.

**Finite Element Modeling.** Finite element modeling was performed using COMSOL. Each of the studied samples was individually modelled using its exact geometry measured by AFM (AC mode). The tip was modelled as a spherical object of radius 25 nm which corresponds to the shape and radius of the tip as determined by scanning electron microscopy (Supplementary Section 2). Simulations performed on similar membranes and for tip radius in the range of 10 nm to 35 nm, lead to equal outcomes (Supplementary Section 6). MoS$_2$ membranes are described by a Young's modulus of E = 270 GPa[10] and Poisson's ratio ν = 0.27.[45]

## ACKNOWLEDGEMENTS

The authors acknowledge S. Bertolazzi and D. Ovchinnikov (EPFL) for valuable discussions. Device fabrication was carried out in the EPFL Center for Micro/Nanotechnology (CMI). We thank Z. Benes (CMI) for technical support with e-beam lithography. This work was financially supported by funding from the European Union's Seventh Framework Programme FP7/2007-2013 under Grant Agreement No. 318804 (SNM) and was carried out in frames of the Marie Curie ITN network "MoWSeS" (grant no. 317451). We acknowledge funding by the EC under the Graphene Flagship (grant agreement no. 604391).

**Author contributions**

SM performed the device fabrication, measurements, data analysis and finite element modelling. AA and AG contributed to finite element modelling. AK designed the experiment, initiated and supervised the work. SM, AA and AK wrote the manuscript.

**SUPPORTING INFORMATION AVAILABLE**

Supplementary figures and discussion related to AFM calibration, additional devices, FEM modeling and piezotronic effects. This material is available free of charge via the Internet at http://pubs.acs.org.

The authors declare no competing financial interest.

**FIGURES**

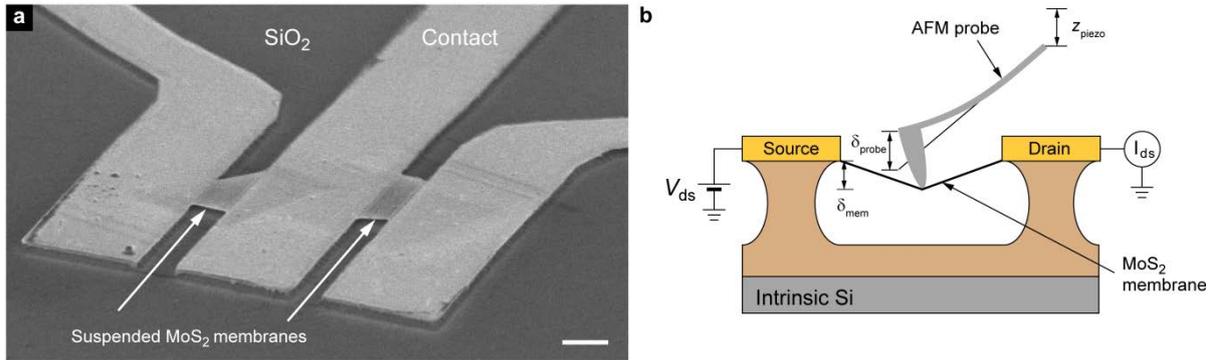

Figure 1. Suspended MoS$_2$ devices and the measurement setup. **a,** Scanning electron microscope (SEM) image of typical MoS$_2$ devices with suspended channels and contact electrodes. Scale bar is 1µm. **b,** Schematic drawing of the of suspended channel MoS$_2$ devices under strain. The suspended atomically thin membrane is deformed at the center using an AFM probe attached to a piezo scanner. The vertical displacement of the scanner ($z_{piezo}$) results in the deflection of the cantilever ($\delta_{probe}$) and the membrane ($\delta_{mem}$). The device is kept under bias voltage $V_{ds}$ while the drain current $I_{ds}$ is monitored.

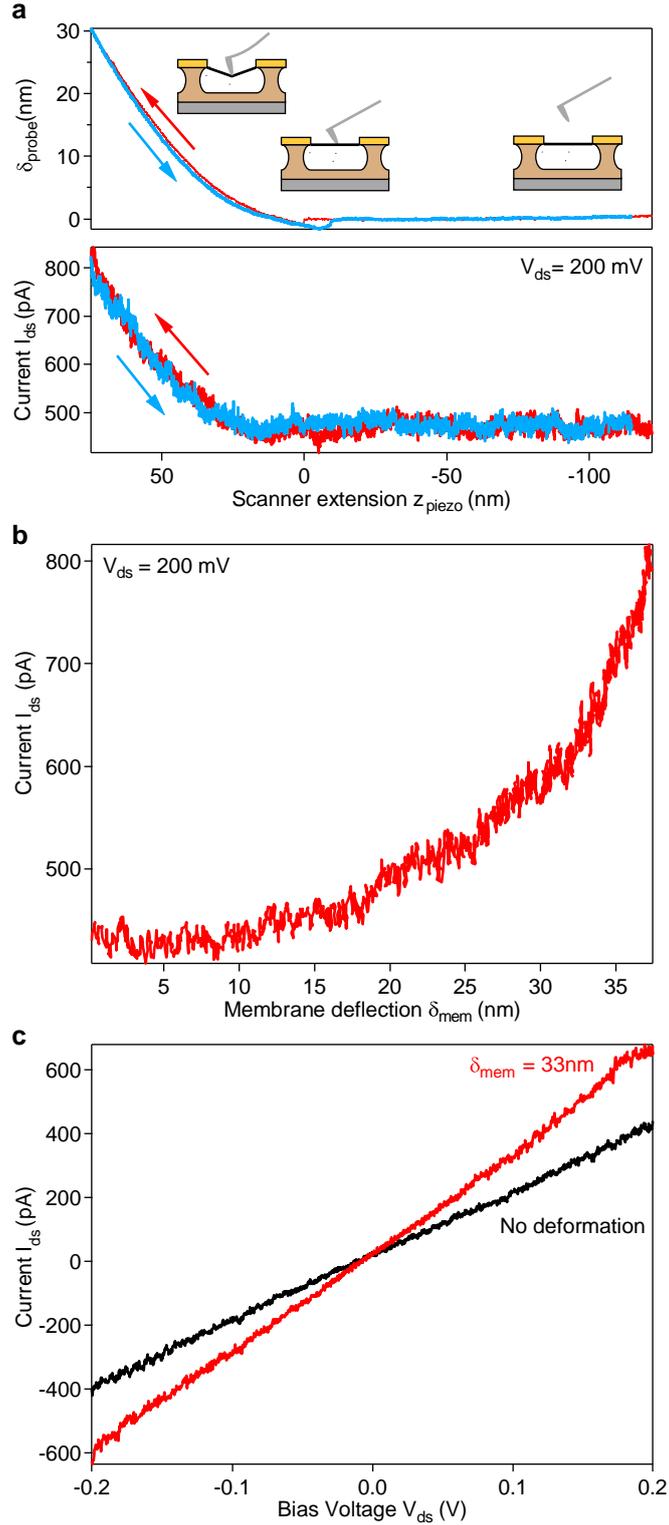

Figure 2. Piezoresistive response of a monolayer MoS$_2$ nanoelectromechanical device. **a,** The output data from electromechanical experiment on device N°1 shows simultaneous measurement of the cantilever deflection (top panel) and the drain current (bottom panel) as a function of the piezo scanner extension. The electromechanical response is reproducible in both extension (red) and retraction (blue) cycles. **b,** Drain current as a function of membrane deflection at the center $\delta_{mem} = (z_{piezo} - \delta_{probe})$. **c,** The output characteristics of the same MoS$_2$ device. Black curve is recorded after the AFM tip has touched the membrane and before indentation. The red curve is recorded while the membrane is kept at constant deformation. The modulation of carrier transport under strain is consistent with the extension and retraction experiments in part a and the piezoresistive effect.

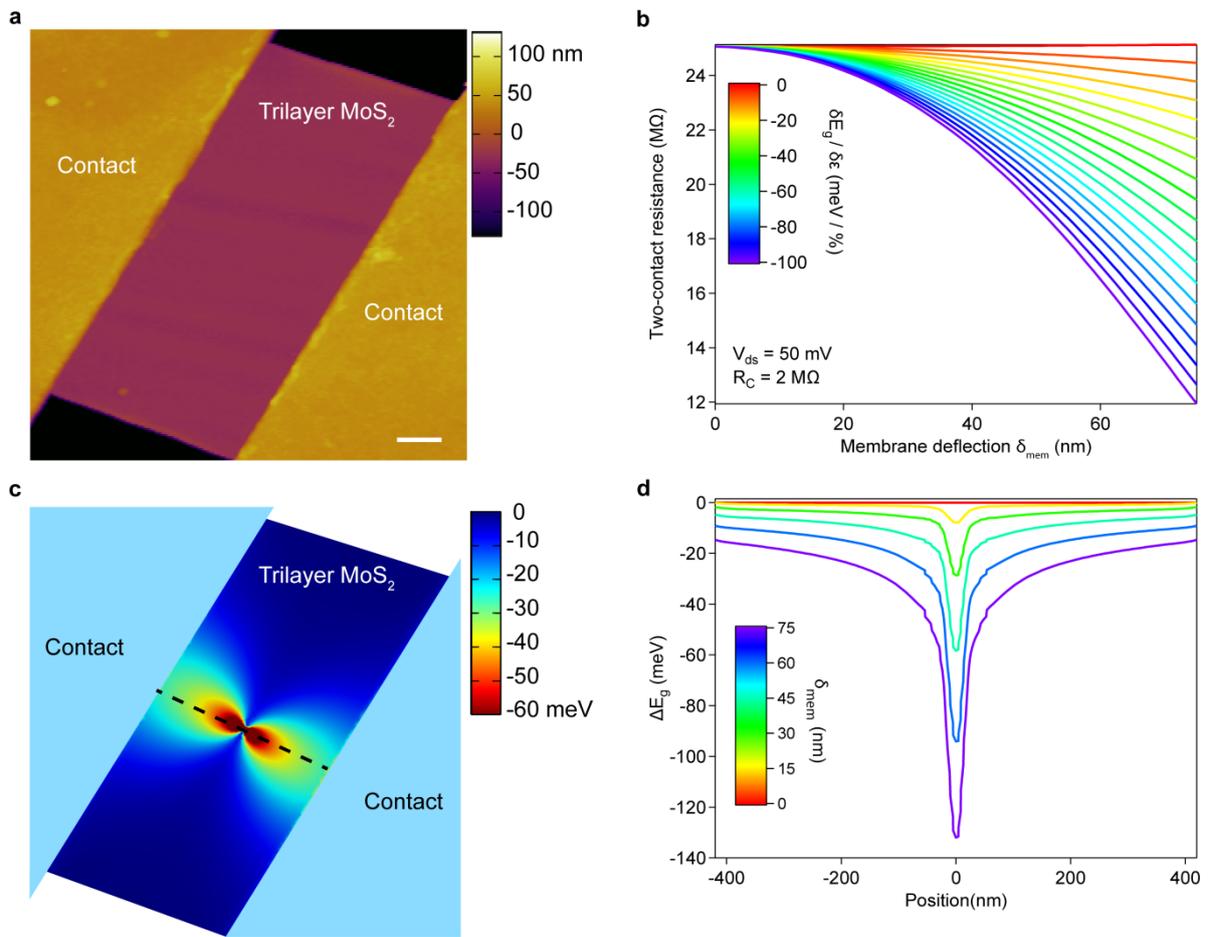

Figure 3. FEM of electromechanical response in a MoS$_2$ membrane under strain. **a,** Topographic AFM image of a trilayer MoS$_2$ suspended membrane clamped with contact electrodes. The scale bar is 200 nm **b,** FEM simulation of the electromechanical response of the MoS$_2$ membrane for different values of $\partial E_g/\partial \varepsilon$ **c,** Simulation result showing the spatial distribution of $\Delta E_g$ under deformation $\delta_{mem}$ = 75 nm **d,** Profile of bandgap change $\Delta E_g$ along the dashed line in c. Areas closer to the tip experience more deformation and thus a higher change in the bandgap.

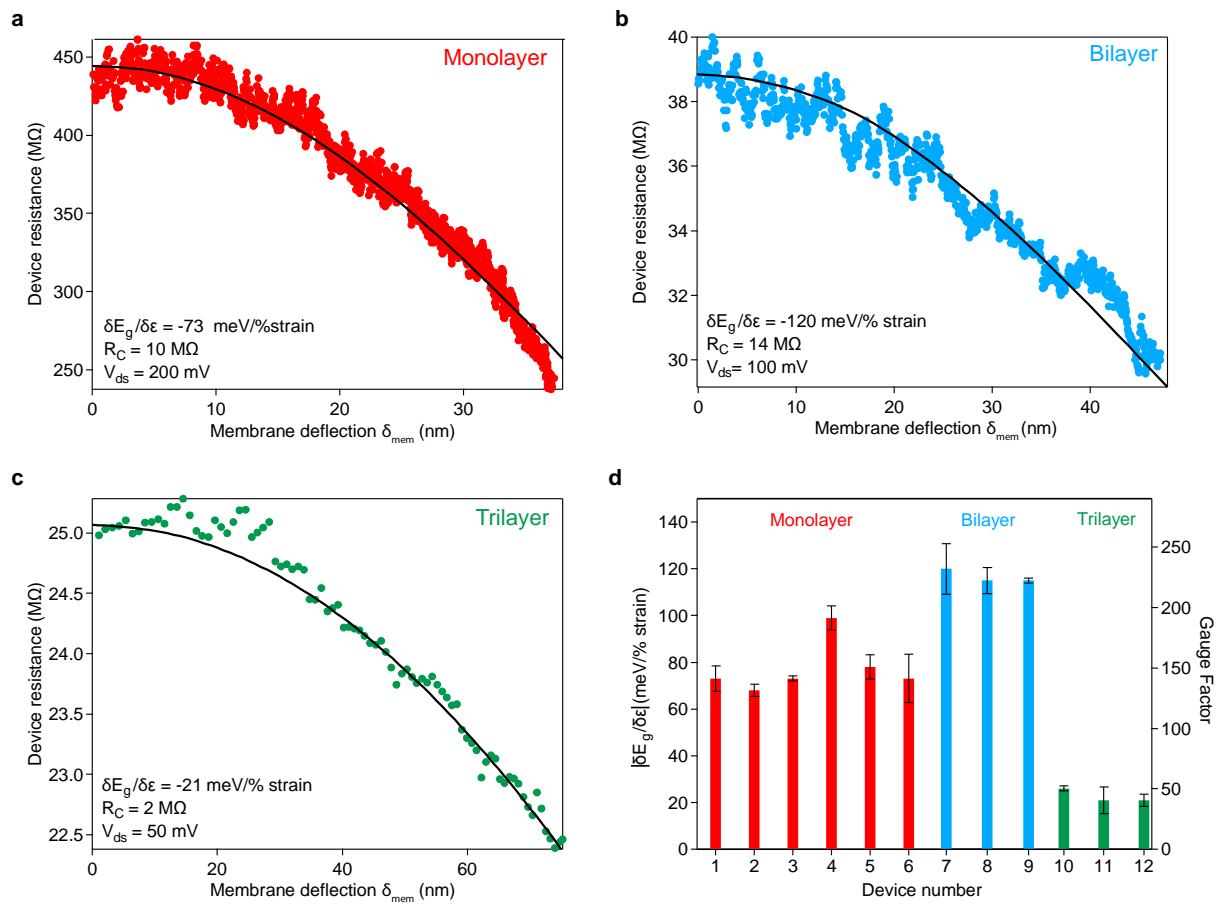

Figure 4. Strain-induced band gap tuning in mono and few-layer MoS$_2$ and modulation of the device resistance due to mechanical deformation of the MoS$_2$ membrane. **a**, Measurements and the corresponding simulation results for monolayer MoS$_2$ indicating a reduction of the band gap $|\partial E_g/\partial \varepsilon|$ with a rate of -73 meV/% **b,** bilayer MoS$_2$ with $\partial E_g/\partial \varepsilon$ = -120 meV/% and **c,** trilayer MoS$_2$ with $\partial E_g/\partial \varepsilon$ = -21 meV/% **d,** Extracted rate of band gap change $|\partial E_g/\partial \varepsilon|$ and piezoresistive gauge factor for 6 monolayers, 3 bilayers and 3 trilayers (for the data on geometry of each device see Supplementary Section 3).